# Complex Evaluation of Hierarchically-Network Systems

**Dmytro Polishchuk, Olexandr Polishchuk, Mykhailo Yadzhak**

Department of Nonlinear Mathematical Analysis, Pidstryhach Institute for Applied Problems of Mechanics and Mathematics National Academy of Sciences of Ukraine, Lviv, Ukraine

*Corresponding author: od_polishchuk@ukr.net

**Abstract.** Methods of complex evaluation based on local, forecasting, aggregated, and interactive estimation of the state, function quality, and interaction of complex system's objects on the all hierarchical levels is proposed. Examples of analysis of the structural elements of railway transport system are used for illustration of efficiency of proposed approach.

*Keywords*: Complex System, Network, Evaluation, Aggregation, Forecasting

## 1. Introduction

Complex system (CS) with hierarchically-network structure are used almost in all areas of human activity, e.g. in transportation (railway, road and aviation systems, transportation networks of large cities and regions of countries), supply and logistics (systems for power, gas, petrol, heat and water supply, trade networks), information and communication (Internet, TV, radio, post service, press, fixed and mobile telephony), in economics (networks of state-owned and (or) private companies, their suppliers and final products distributors), finance (banking and insurance networks, money transfer systems), education, healthcare etc. Their state and functioning quality impose large impact on citizens' quality of life, efficiency of economy and possibilities for its development, readiness of government structures to mitigate impacts of technological and natural disasters. Finally, they may be treated as the evidences of country development level in general. All these circumstances determine the relevance of development of methods for CS study. Solution for this problem belongs to the fields of systems theory, system analysis, complex networks theory, mathematical modelling etc [1-3]. Complex systems appear, operate and develop within long periods of time and with natural processes of "aging", despite regular improvements, more strict and accurate control over their behaviour is required. This is why the development of methods for evaluation and forecasting the state, quality of functioning and inter-action between structural elements of CS, presented in this work, takes especially important place.

Methods of evaluation can be based on deterministic [4, 5], statistical [6, 7], stochastic [8] or hybrid [9] approaches. Each of them has its own benefits and disadvantages [10]. Deterministic methods generate an evaluation of the real state and functioning quality of a particular object in the system [11, 12]. However, a careful analysis of all objects in the system is in many cases impossible. Then use the statistical and stochastic methods [13-15]. They provide an opportunity to pay attention to the basic problems functioning of the system, for example, the most common diseases of the population in the region [8]. However, these methods often do not allow us to determine objects, malfunctions which could lead to crashes of separate subsystems or system on the whole.

The most of methods of evaluation focus on procedures of aggregation [10, 16, 17]. However, fuzzy or unilateral local evaluations do not allow us to form reasoned generalized conclusion and build an accurate forecast of behavior of the system even on a short-term perspective. Multi-criteria and multi-parameter analysis of the functioning of the elements of the system is the basis for the formation of objective quantitative evaluations at all levels of the hierarchy. Aggregation neglects both positive and negative results of the local evaluation [5]. Therefore, the procedures of generalization should be accompanied by a means of top-down analysis of the behavior of system objects [18, 19].

On functioning of the real system is influenced by many internal and external factors. They can create risks that can not be foreseen by regular scheduled investigations. Therefore, special attention should be paid to the continuous monitoring of processes occurring in the system [20-22].

In general, the methods used to evaluate the state and functioning quality of the system should focus on the type, structure, function and its other features. Sufficiently detailed overview of the methods of evaluation and the peculiarities of their use can be found in [4, 6, 10, 14, 23].

Theory of evaluation of complex systems is a component of the system analysis [19]. On the other hand, evaluation results are objective and the most significant reason for making an informed decision regarding further action on the studied system [8, 25]. In this information content of evaluation, its understandability, and convenient procedures for operational orientation in a large number of obtained conclusions allow to make timely organizational and management decisions. Note that the conclusions drawn on the basis of deterministic estimation of systems generate a much smaller amount of alternatives than statistical or stochastic.

## 2. Complex Hierarchically-Network Systems

Existence of complex systems of different types, destinations and structures operating under different rules and conditions has initiated the number of system's definitions, none of which has become commonly accepted.

This is why, at the present stage, the approach presented by M.P. Buslenko [3] appears to be the most acceptable, where main characteristics defining certain object as a system are presented. They include, in particular: presence of certain number of interconnected elements, functions they perform and directions for reaching defined purposes of their functioning; ability of system to be split into subsystems the functional purposes of which are subordinated to the overall goal of the system; presence of control and extended informational network, intensive material and informational flows, interaction with external environment and ability to operate under presence of random factors. Systems comprising dozens of thousands of elements are referred to as large [24]. CS is considered dynamic if its state changes with time [3]. System complexity is quite a relative concept [26], in particular, the more levels of system's splitting into subsystems there exist, the more objects constitute those subsystems, the higher the diversity of such objects is, the more functions they perform, the more interaction with other objects they implement, the more ways to react on the action of internal and external factors there is possible and the higher the diversity of such reactions is, the more complex is the system in general.

Study of system usually begins with definition of its components and structure. The most widespread types of structure for existing complex systems include hierarchical, network and hybrid, in particular, hierarchically-network. The peculiarity of hierarchically-network structure (HNS) is presented by fact that depending on purpose and depth of study on the level of elements or the subsystems of the lowest level of splitting, which hereinafter will be referred to as basic, they are the collection of nodes, connected by edges through which the flows are passing. The edges, therefore, shall ensure smooth passage of flow and nodes are to ensure its processing. Hierarchy is introduced on the basis of management system construction principles, CS objects arrangement in space etc. The notion "object of system" will hereinafter designate structural unit of system of arbitrary hierarchy level – from subsystem element to highest level of splitting.

The example of complex hierarchically-network system (CHNS) we shall hereinafter use to illustrate proposed evaluation methods is railway transport system (RTS) of country. The structure of the railways in most countries includes thousands of stations, deployed length of railway lines in the tens of thousands of kilometres, the number of locomotives, freight and passenger cars exceeds hundreds of thousands of units. Their activities provide hundreds of enterprises (depots, track machine stations, power supply, alarm and communication sections) and hundreds of thousands of employees. In many countries, national railways provide more than 50% of passenger and freight traffic.

As for RTS, the determinative sign of its consecutive splitting into subsystems of lower levels is strict territorial hierarchical principle of national railways management system framework. For example, Ukrainian railways includes 6 territorial railways, 27 directories of railway transportation, 110 track distances and over 1200 divisions that usually are represented by sequence of stations and inter-station railway tracks with the approximate duration of 20–30 km [27]. Such structuring principle allows us to determine clear connection between RTS objects and sub-units of national railways responsible for their state and functioning quality. Further structuring is implemented according to functions performed, in particular, basic subsystems (BSSs) composing divisions are represented by following objects: stations (nodes) and inter-station railway tracks (edges). Trains are representing flows in such system. BSSs also can be complex systems, for example, junction stations and stations of large cities. It is reasonable to study such systems separately. Another peculiarity of HNS consist in fact that on every higher level of hierarchy it is network, the flows for which being information, organizational and management decisions etc.

When splitting basic subsystems (BSS) into elements we take into account following considerations. Every system implements certain set of functions out of which, depending on purpose of study, the main function is selected. Determination of elements is performed from the point of view of their participation in main function implementation. Elements not participating in implementation of this function are not included into system components or its structural schema in the process of evaluation. Thus, the main function of RTS is to provide reliable and safe train movement in accordance with schedule established. This determines the composition of objects in the system, subject to evaluation.

BSS may be split in to the elements of one or several types. There is set of characteristics corresponding to every element describing its state and functioning process. Every characteristic has corresponding range of permissible values, limitations in time and space etc. Thus, if inter-station railway track is considered as BSS of RTS, it is reasonable to split such track into elementary sections divided by artificial objects (bridges, crossings etc.) that differ in space, structural, geological features etc. and their length does not exceed e.g. 1 km. Such division is explained, in particular, by fact that, on horizontal line, rails of track shall lie on the same altitude, while on the curve external rail shall lay few centimetres (depending on curvature of track) higher. This is also relevant for roadbed geometry. It is obvious that, when studied, such sections shall be considered separately, since the range of permissible values of track characteristics for them differ greatly. Thus, negative values for characteristics, for instance, location of tracks in space, are different for straight, curve or inclined sections. This difference is so significant that permissible values of one type of section are not valid for another one and vice versa.

Main components of station are its railroad and other objects of station infrastructure involved in providing of main function of station, i.e. smooth train passage or its acceptance, maintenance and dispatch. Their splitting into elements is performed according to the above described principle.

Similar structuring method is applied to most CHNS, for instance, road transport network of the country, region or district. Nodes in such network are represented by inhabited localities, edges – by highways connecting them, flows – by motor vehicles. Such a method not only makes the analysis of system simpler, it also establishes direct

It is usually difficult to implement classical mathematical modelling methods [3] on practice for studying most existing CHNS due to the problems of dimension and adequacy. Network analysis methods [2] are focused mostly on studying interconnections between network BSSs without analysis of their elements' state and functioning quality. At the same time, flow processing in the node may be quite a complicated process as well, for example it will include acceptance, reconfiguring, loading and detraining of load, state inspection, dispatching freight train from the station etc. Evaluation of existing CHNS requires system approach, which we propose to implement within the methods of complex evaluation (MCE) of the system.

## 3. Methods for Evaluation of Complex Systems

Usually, two main approaches are applied to control state and behaviour of existing CHNS: regular scheduled inspections, distinctive features of which are accuracy and possibility for further development of recommendations for elimination of drawbacks discovered; and continuous monitoring of system objects' functioning that allows us to draw mediate, but still significant conclusions regarding its actual state and functioning quality. Thus, at railways, the evaluation of state of track is performed on regular basis at least twice a year and for its monitoring following up the trains movement may be used with an aim to define whether it corresponds to time-schedule established [27].

It is reasonable to start evaluation of real systems with objects of lowest structural level, i.e. with elements of BSSs. We define an element as an object of clearly defined location, functional destination and relevant set of characteristics describing its state and functioning process with corresponding ranges of permissible values for those characteristics. All characteristics are evaluated according to certain collection of criteria and parameters. Of course, evaluation of every object presupposes evaluation of its state on the first place, and only after that the evaluation of quality of implementation of its functions that in any case depend on element's state – either directly or indirectly. The process of evaluation is started only after the stage of thorough selection and processing of experimental data as to each of characteristic and their conversation into format, suitable for further analysis. Thus, data regarding the state of rails that allow us to discover cracks are collected by means of defectoscope with the step of 1mm. Considering the maximal length of elementary section, which comprises 1 km, this means that the data array of the size of 1 million values may be created. It is obvious, that to be used for adequate analysis such data require relevant processing and formatting.

Currently, for evaluation of CHNS, and RTS in particular, integer rating or conceptual ("excellent", "good", "satisfactory", "unsatisfactory") scale [20] is commonly used. Its main drawback is that "satisfactory" evaluation may imply wide range of concepts – from "almost good" to "slightly better than unsatisfactory". We propose unified approach for evaluating state, quality of functioning and interaction between system structural elements, which consists in developing main rating evaluation and its adjustment with regard to type and features of object studied. Such an approach allows not only to compose more clear understanding of evaluated object, but also to localize the reasons for drawbacks discovered.

The number of characteristics describing BSS may comprise dozens [12]. Different characteristic may be selected in different ways and they priority regarding structure and functions of element may be different. It is clear that the conclusions as to separate characteristics are to be generalized with consideration of their priority. Recording the number of actually evaluated elements' characteristics is also important. From this point on, evaluations for elements' state and functions they implement on the basis of their characteristics behaviour analysis will be referred to as *local*. In some cases it is reasonable to limit local evaluation with BSS level of system without excessive detail of their components.

As usual, scheduled inspections of system's objects are held at different time points, which means the results of last study may not stay on such stage till following inspection, and state of object and its functioning quality may cross "safety threshold" [28]. It should be also taken into account that every real system evolves in time, i.e. with regard to current requirements, its evaluation may be insufficient. Therefore, evaluation process should contain means of analysis of CS's meeting expected requirements for short- and long-term perspective. Thus, the evaluation process should not only determine conclusions and discover "faulty" elements for the time point moment when study is held, but also it should forecast further behaviour of system objects. Forecasting analysis performed on the basis of local evaluations prehistory, allows us to determine the nature, direction and speed of system state change, follow up negative processes and forecast potential risks, as well as material and financial expenses required for their elimination or timely prevention. Regarding railways, it means that its structural elements are ready to seasonal changes in passenger and freight flow or to radical modernisation of separate subsystems, which is required e.g. for implementation of high-speed railway traffic. Let us refer to above described method as to **forecasting**.

Number of local evaluations of real CHNS may reach dozens of millions values [12], which obviously exceed the capacity of their manual analysis. For their generalization, i. e. for developing conclusions regarding their state, quality of functioning and interaction of objects of higher hierarchy levels (subsystems and CS in general), tools of linear and non-linear aggregation are applied [5], taking into account weighted coefficients that reflect importance of separate objects in system's structure and priority of functions they perform. World practice of transport systems usage provides quite lot of examples of unsatisfactory state or functioning quality of one system object that results in catastrophes with numerous human victims and significant material losses. This is why generalized evaluations for all levels may both arise as the result of weighted averaging and be equal to the evaluation of the "weakest" object of the system. Evaluations of second type are applied in cases when failure of one object constitutes real threat to functioning of e.g. some BSS of system. Since weighted averaging mitigates the results of both positive and negative evaluations, it is reasonable to make generalisation of conclusions after elimination of causes and revaluation of drawbacks eliminated. Let us refer to above described method as to **aggregated.** Surely, together with means of bottom-up analysis, implemented by aggregating procedures, the evaluation process should also contain tools for top-down analysis for localization of objects, results for which appeared to be negative or close to negative.

Due to the number of reasons, scheduled inspections may often not discover drawbacks that arise "out of schedule". It should be also taken into account that even excellent state and functioning quality of separate objects in the system do not ensure high performance of its subsystems or system in general. And vice versa, the most optimal work organization process will not ensure high efficiency of system functioning if CHNS's state or organization of components functioning is unsatisfactory. The more worn-out CHNS's objects are the more urgent is the problem of continuous monitoring of their state and functioning process. Quality of implementation of functions by object may be affected by number of third-party factors, both internal and external as to the system. Internal influence may be evaluated on the level of subsystems connecting interacting objects. We shall call this evaluation method *interactive*. It allows us to determine separate objects in selected subsystem, functioning of which is unsatisfactory, without thorough analysis of state and functioning quality of these objects and expenses related to such analysis. The simplest interactive evaluation may be performed for system where the movement of flows is deterministic, at least partially, in accordance with certain schedule, the compliance to which may be periodically summed up. Railway system belongs to systems of such type, since the railway traffic within it is fully determined. Transportation systems of great cities are partially determined, since they are largely affected by random factors (traffic accident may cause traffic jams, or reallocation of transport flows into alternative roads). However, compliance with traffic schedule by community transport allows us to draw at least mediated conclusions as to effectiveness of city traffic organization. Similar examples of organization of continuous monitoring may be provided for other CHNS. It is reasonable to include generalized results of interactive evaluation over certain time period between two scheduled inspections into aggregated evaluation procedure. Those results may be also used for more detailed and accurate forecasting analysis of functioning of evaluated system's objects.

In general, only if combined, proposed methods may provide sufficiently full and adequate understanding of CHNS quality. Indeed, high local evaluations do not ensure effective interaction of elements, failures of separate systems objects may result in breakdown in balanced organization, satisfactory state of object for the moment of current inspection does not imply the state will stay satisfactory till the next inspection. Huge amount of information regarding separate CHNS elements without appropriate generalization is ill-suited for rapid analysis and timely reaction for drawbacks discovered. On higher generalization levels, evaluation allows to determine reliable conclusion as to the state and functioning quality of system and its main subsystems and to define measures, as well as material and finance expenses required for its modernisation and optimization of functioning. At the local level evaluation allows to identify separate elements and their components subject to improvement. These "narrow" places that are constantly discovered during scheduled inspections or continuous system monitoring may be subject to mathematical modelling. This narrows down the object of modelling and makes the process itself more realistic.

Significant volumes of information received during evaluation process require development of such means of their visualization that would ensure quick orientation in large number of outcomes received. This is why together with development of evaluation algorithms it is also important to elaborate principles of their implementation results visualisation and instruments of bottom-up and top-down analysis of such results.

In general, collection of above described interconnected methods and approaches comprises the methods for complex evaluation of system, the schema for which is presented on the fig.1. It defines the way for reflecting CHNS experimental studies data onto structured, according to hierarchy, sequence of local, forecasting, interactive and generalized evaluations for their state, functioning and interaction with other system objects quality. Taking into account the diversity of CHNS objects, MCE defines universal principles of such evaluations development, common for all objects of the same type and functional destination considering peculiarities of the former. Let us describe main MCE components in details.

## 4. Local Evaluation

Local evaluation of complex systems elements is defined by type of characteristics describing them and appearance of reference ranges and ranges of permissible values for these characteristics. Let us consider few most common cases.

Let us perform the evaluation of characteristic $f(x)$, $x \in [0, X]$, where $x$ is spatial or time variable, describing state or functioning process of system element, according to the following algorithm. Let us assume that $F[0, X]$ is range of permissible values of $f(x)$, $x \in [0, X]$, and $\{F_i[0, X]\}_{i=3}^{5}$ are subranges of $F[0, X]$, defining main positive integer rating evaluation $e(f)$ of characteristic $f(x)$. I. e. evaluation $e(f) = i$ if $f \in F_i[0, X], i = 3, 4, 5$, and $e(f) = 2$ if $f \notin F[0, X]$. Adjusted rating evaluation is developed as follows:

$$E(f) = e(f) + (1 - \|P_{F_i}(f)\|_V / v_i),$$

in case $e(f) = i$, $i=3,4$. Here $P_{F_i}(f)$ is projection onto subrange $F_i[0, X]$ of values of characteristic $f(x)$, main rating evaluation for which is equal to $i$, $v_i$ is normalizing coefficient, $\|.\|_V$ is norm of functional space $V$. For instance, for $C_0[0, X]$ the value $v_i = \max_{x \in F_i[0,X]} f(x) - \min_{x \in F_i[0,X]} f(x)$, and in case $L_2[0, X]$ the value $v_i$ is proportional to area of subrange $F_i[0, X]$, $i = 3,4,5$. In the case then reference subrange $F_5[0, X]$ degenerates into the curve $f_{ref}(x)$, proposed algorithm evaluates measure of deviation of characteristic $f(x)$ from its accepted reference behaviour or from solution of corresponding optimal control problem, in case it is possible to find one etc. This algorithm is also used for evaluation of first derivative of characteristic $f(x)$ and provides the possibility to analyse dynamics of its variation in interval $[0, X]$. Thus, oscillating dynamics of state of track characteristics within section indicates decrease in comfort and safety of railway traffic, especially when train speed increases.

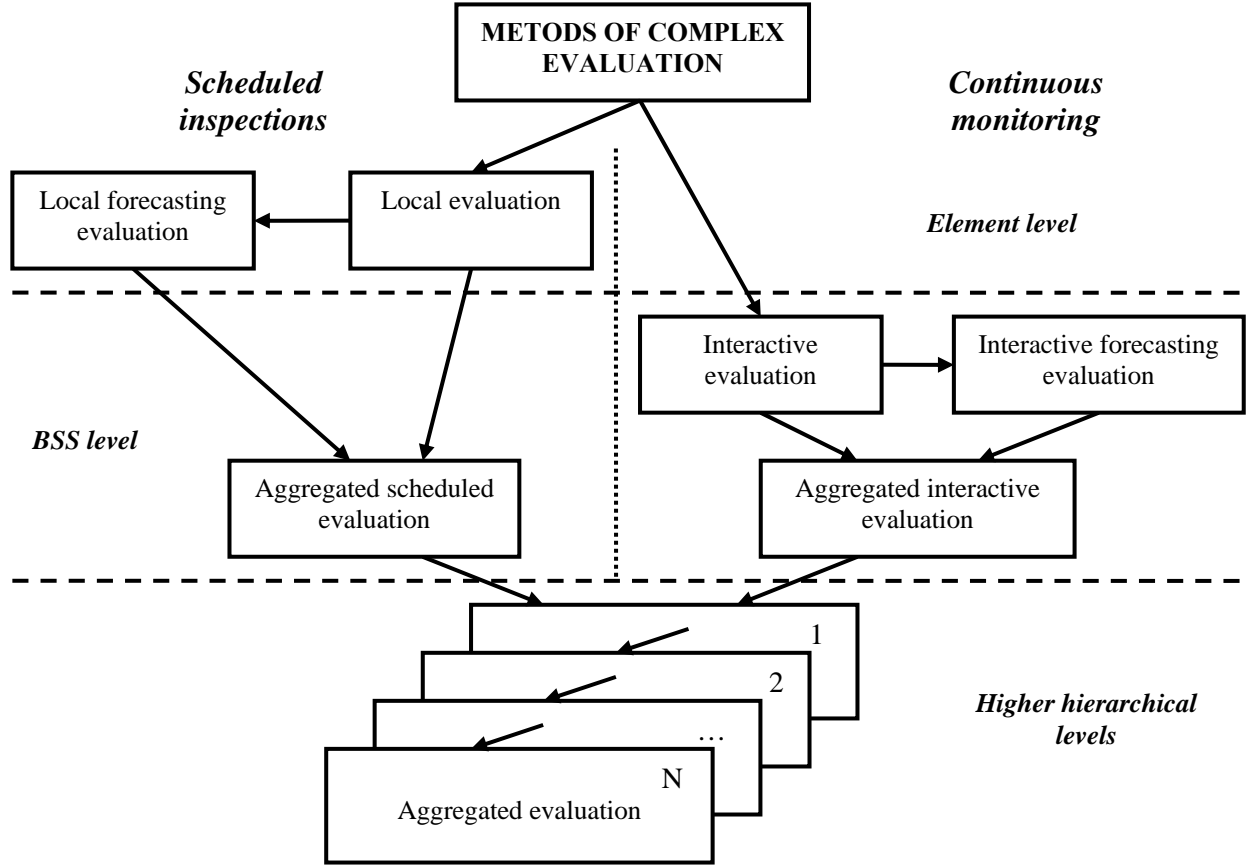

**Fig.1.** Schema of methods of complex evaluation of system

Let us consider the algorithm for local evaluation for case $F_5[0, X] = f_{ref} \equiv const$ in detail. Let us suppose that $SD = \{S_m\}_{m=1}^{\tilde{M}} \cup \{D_m\}_{m=1}^M$ is subsystem composed from interconnected BSS, i. e. it is the collection of nodes $S_m$, $m = \overline{1, \tilde{M}}$ and edges $D_m$, $m = \overline{1, M}$, connecting them. In case of RTS the simplest example of such system is division in the form of sequence of stations and tracks between them. Let us divide the edge $D_m$ into the sequence of elementary sections $\{D_{m,n}\}_{n=1}^{N_m}$ of the length $X_{m,n}$, state of each being described by the set of characteristics $\{f_{m,n,i}(x)\}_{i=1}^{I_{N_m}}$, $x \in [0, X_{m,n}]$, $n = \overline{1, N_m}$. Let us suppose, that for characteristic $f_{m,n,i}(x)$ range of permissible values is presented with $F_{m,n,i}[0, X_{m,n}] = \{f_{m,n,i}(x): f_{m,n,i}^{ref} \leq f_{m,n,i}(x) \leq f_{m,n,i}^{\max}, x \in [0, X_{m,n}]\}$, where $f_{m,n,i}^{\max}$ is maximal possible deviation of characteristic $f_{m,n,i}(x)$ from its reference value $f_{m,n,i}^{ref}$. Let us define subranges $F_{m,n,i}[0, X_{m,n}]$ for characteristic $f_{m,n,i}(x)$ behaviour. Subranges correspond to different values of integer rating evaluation scale. It is considered "excellent" in case $f_{m,n,i}(x) \equiv f_{m,n,i}^{ref}$. Subrange ($f_{m,n,i}^{ref}$, $\gamma$] corresponds to "good" behaviour. Values $\gamma \in [f_{m,n,i}^{ref}, f_{m,n,i}^{\max}]$ are set by experts. Subrange $(\gamma, f_{m,n,i}^{\max}]$ corresponds to "satisfactory" behaviour. If the value of characteristic $f_{m,n,i}(x)$ in any point of interval $[0, X_{m,n}]$ exceeds the value $f_{m,n,i}^{\max}$, its behaviour is considered "unsatisfactory". It is obvious that behaviour of characteristic $f_{m,n,i}(x)$ is defined by its greatest deviation from reference, i. e. by value

$$\left\| f_{m,n,i}(x) - f_{m,n,i}^{ref} \right\|_{C_0[0, X_{m,n}]} = \max_{x \in [0, X_{m,n}]} \left| f_{m,n,i}(x) - f_{m,n,i}^{ref} \right|.$$

The example of such characteristic for elementary section of track, that constitutes horizontal line, is characteristic defining mutual position of rails in vertical plane when train is moving with maximal weight and speed possible for current section.

Evaluation of behaviour of studied characteristic will be performed according to two parameters. According to the first parameter, adjusted rating evaluation $e_c(f_{m,n,i})$ will be made on the basis of value of vertical disturbances $f_{m,n,i}(x)$ on elementary section $D_{m,n}$, in particular, we shall consider $e_c(f_{m,n,i})$ equal to:

– 2, if $f_{m,n,i}^{\max} < \left\| f_{m,n,i}(x) \right\|_{C_0[0, X_{m,n}]}$ ;

– 3 + $(f_{m,n,i}^{\max} - \left\| f_{m,n,i}(x) \right\|_{C_0[0, X_{m,n}]})/(f_{m,n,i}^{\max} - \gamma)$,

  if $\gamma < \left\| f_{m,n,i}(x) \right\|_{C_0[0, X_{m,n}]} \leq f_{m,n,i}^{\max}$ ;

– 4 + $(\gamma - \left\| f_{m,n,i}(x) \right\|_{C_0[0, X_{m,n}]})/\gamma$,

  if $f_{m,n,i}^{\max} < \left\| f_{m,n,i}(x) \right\|_{C_0[0, X_{m,n}]} \leq \gamma$ ;

– 5, if $\left\| f_{m,n,i}(x) \right\|_{C_0[0, X_{m,n}]} \equiv f_{m,n,i}^{ref}$.

According to the second parameter, the adjusted rating evaluation $e_l(f_{m,n,i})$ will be made on the basis of analysis of mass disturbances $f_{m,n,i}(x)$ on elementary section $D_{m,n}$, in particular, we shall consider $e_l(f_{m,n,i})$ equal to:

– 2, if $f_{m,n,i}^{\max} < \left\| f_{m,n,i}(x) \right\|_{C_0[0, X_{m,n}]}$ ;

– 3 + $((f_{m,n,i}^{\max} - \gamma)\sqrt{X_{m,n}} - \left\| f_{m,n,i}(x) - \gamma \right\|_{L_2[0, X_{m,n}]})/$

  $/(f_{m,n,i}^{\max} - \gamma)\sqrt{X_{m,n}}$, if

  $\gamma < \left\| f_{m,n,i}(x) \right\|_{C_0[0, X_{m,n}]} \leq f_{m,n,i}^{\max}$ ;

– 4 + $(\left\| \gamma - f_{m,n,i}(x) \right\|_{L_2[0, X_{m,n}]})/\gamma \sqrt{X_{m,n}}$ ,

  if $f_{m,n,i}^{\max} < \left\| f_{m,n,i}(x) \right\|_{C_0[0, X_{m,n}]} \leq \gamma$ ;

– 5, if $\left\| f_{m,n,i}(x) \right\|_{C_0[0, X_{m,n}]} \equiv f_{m,n,i}^{ref}$.

Consider the examples of behavior of characteristics for which the main integer evaluation is "satisfactory" (Fig. 2). Evaluations of characteristics displayed lines 1 and 2 in the uniform metric are equal namely $e_c(f_{m,n,1}) = = e_c(f_{m,n,2}) = 3.10$. However, their behavior differs significantly. If the values of the first characteristic are close to the critical at only one point, the second - on the whole interval. Similarly, the evaluations of characteristics displayed lines 1 and 3 in the mean-square metric are equal namely $e_l(f_{m,n,1}) = e_l(f_{m,n,3}) = 3.94$. However, if the first of these characteristics has a critical point, the second is "almost good" on the whole interval. Thus, the use of one-parameter evaluations do not provide an adequate understanding of the behavior of the investigated system elements characteristic.

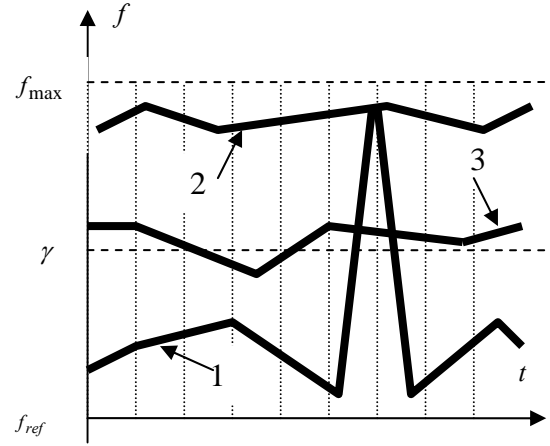

**Fig. 2.** Behaviour of characteristic

At the same time the pair $e_c(f_{m,n,1}) = 3.10$, $e_l(f_{m,n,1}) = 3.94$ means that within selected elementary section there are only points where values $f_{m,n,i}(x)$ are close to acceptable limits and they can be eliminated during simple local improvement (Fig. 2, line 1). Pair of evaluations $e_c(f_{m,n,2}) = 3.10$, $e_l(f_{m,n,2}) = 3.12$ indicates that state of section with regard to characteristic studied is close to critical and requires serious improvement (Fig. 2, line 2). Pair of evaluations $e_c(f_{m,n,3}) = 3.95$, $e_l(f_{m,n,3}) = 3.94$ designates that state of section with regard to characteristic studied is close to "good" (Fig. 2, line 3). I. e., developed evaluations provide quite specific, reasonable and understandable for average user information, for example, considering state of track study: local disturbances, that can be eliminated through simple repair in points, in general almost good state, state that requires urgent track repair etc.

Element's state in case of absence of unsatisfactory evaluations of its characteristics in uniform and mean-square metrics is evaluated according to formulas

$$H_{m,n}^c = <\boldsymbol{\rho}_m, \mathbf{e}^c(\mathbf{f}_{m,n})>_{R^{I_{N_m}}} / <\boldsymbol{\rho}_m, \mathbf{1}>_{R^{I_{N_m}}} \quad (1)$$

$$H_{m,n}^l = <\boldsymbol{\rho}_m, \mathbf{e}^l(\mathbf{f}_{m,n})>_{R^{I_{N_m}}} / <\boldsymbol{\rho}_m, \mathbf{1}>_{R^{I_{N_m}}} \quad (2)$$

accordingly. Hereinafter the $<.,.>_{R^K}$ is scalar product in Euclidean space $R^K$, $\mathbf{e}^c(\mathbf{f}_{m,n}) = \{e^c(f_{m,n,i})\}_{i=1}^{I_{N_m}}$, $\mathbf{e}^l(\mathbf{f}_{m,n}) = \{e^l(f_{m,n,i})\}_{i=1}^{I_{N_m}}$ are vectors of adjusted rating evaluations for characteristic $f_{m,n,i}$ of element $n$ of basic subsystem $D_m$ in uniform and mean-square metrics, $\boldsymbol{\rho}_m = \{\rho_{m,i}\}_{i=1}^{I_{N_m}}$ is the vector of weighted coefficients defining the priority of element characteristics, $\mathbf{1} = \{1\}_{l=1}^{I_{N_m}}$, $n = \overline{1, N_m}$, $m = \overline{1, M}$.

Generalized conclusion about the state of elementary section $D_{m,n}$ with regard to the set of evaluation parameters is defined according to formula

$$H_{m,n} = (\rho^c H_{m,n}^c + \rho^l H_{m,n}^l)/(\rho^c + \rho^l) \quad (3)$$

where $\rho^c$ and $\rho^l$ are weighted coefficients defining priority of evaluation parameters, $n = \overline{1, N_m}$, $m = \overline{1, M}$. Similarly is evaluation of the characteristics of elements of nodes $S_m$, $m = \overline{1, \widetilde{M}}$, of subsystem $SD$ performed.

Another important evaluation indicator is the level of evaluation coverage of characteristics of BSS element, with regard to their priority, that is defined according to formula

$$C_{D_{m,n}} = 100\% \times <\boldsymbol{\rho}_m, \boldsymbol{\delta}_{m,n}>_{R^{I_{N_m}}} / <\boldsymbol{\rho}_m, \mathbf{1}>_{R^{I_{N_m}}} \quad (4)$$

where $\boldsymbol{\delta}_{m,n} = \{\delta_{m,n,i}\}_{i=1}^{I_{N_m}}$,

$$\delta_{m,n,i} = \begin{cases} 0, & \text{if evaluation of } i-\text{th} \\ & \text{characteristic has not been held}, \\ 1, & \text{if evaluation of } i-\text{th} \\ & \text{characteristic has been held} \end{cases}$$

There are lot of examples of CHNSs, state and functioning of objects of which may be characterized not only by functional relations, on which above described local evaluation algorithms were mainly focused, but also by set of numerical values reflecting parameters of state or certain functions implementation. Such characteristics include, in particular, time intervals for implementation of operations with flow in the node, for instance, with train on the station. Evaluations for such processes are developed on the basis of analysis of numerical deviations of actual operation carrying out from that one established in schedule.

Let us suppose that $\mathbf{O} = \{O_n\}_{n=1}^N$ is complete cycle of operations that need to be carried out over the flow $P$ in node $S$, $\mathbf{O}_{N_i} = \{O_{n_1}, O_{n_2}, ..., O_{N_i}\} \subset \mathbf{O}$, $\sum_{i=1}^I N_i = N$, are the subsets of operations that need to be carried out in sequence, moreover, operations from different subsets $\mathbf{O}_{N_i}$, $i = \overline{1, I}$, may be carried out simultaneously. Let us denote $\boldsymbol{\tau}^s = \{\tau^s(O_n)\}_{n=1}^N$ where $\tau^s(O_n)$ is the time for which operation $O_n$ is carried out according to schedule, $\boldsymbol{\tau}^{\min} = \{\tau^{\min}(O_n)\}_{n=1}^N$, where $\tau^{\min}(O_n)$ is the minimal possible time for which operation $O_n$ is carried out, $\boldsymbol{\tau}^r = \{\tau^r(O_n)\}_{n=1}^N$, where $\tau^r(O_n)$ is real time for which operation $O_n$ is carried out. Evaluation of quality of node functioning should be performed regardless to previous flow delays that occurred for reasons beyond control of current node.

Let us denote

$$\tau^{\min}(\mathbf{O}_{N_i}) = \sum_{n_1, n_2, ..., N_i} \tau^{\min}(O_n),$$

$$\tau^s(\mathbf{O}_{N_i}) = \sum_{n_1, n_2, ..., N_i} \tau^s(O_n),$$

$$\tau^r(\mathbf{O}_{N_i}) = \sum_{n_1, n_2, ..., N_i} \tau^r(O_n).$$

We shall consider that the quality $e_S(P, \mathbf{O}_{N_i})$ of carrying out the subset of operations $\mathbf{O}_{N_i}$ for real processing of flow $P$ in node $S$ is equal to:

– 5, if $\tau^r(\mathbf{O}_{N_i}) = \tau^{\min}(\mathbf{O}_{N_i})$;

– $4 + (\tau^r(\mathbf{O}_{N_i}) - \tau^s(\mathbf{O}_{N_i}))/(\tau^{\min}(\mathbf{O}_{N_i}) - \tau^s(\mathbf{O}_{N_i}))$,

  if $\tau^{\min}(\mathbf{O}_{N_i}) < \tau^r(\mathbf{O}_{N_i}) \leq \tau^s(\mathbf{O}_{N_i})$;

– $3 + (\tau^r(\mathbf{O}_{N_i}) - T_{PS}^s)/(\tau^s(\mathbf{O}_{N_i}) - T_{PS}^s)$,

  if $\tau^s(\mathbf{O}_{N_i}) < \tau^r(\mathbf{O}_{N_i}) \leq T_{PS}^s$;

– 2, if $\tau^r(\mathbf{O}_{N_i}) > T_{PS}^s$, $i = \overline{1, I}$,

where $T_{PS}^s$ is the time interval for which the flow $P$ is in node $S$ according to schedule. The same algorithm will be applied for evaluation of the quality of carrying out $e_S(P, O_n)$ separate operation from sequence $\mathbf{O}_{N_i}$.

Quality $e_S(P)$ of flow $P$ processing in node $S$ is determined according to formula

$$e_S(P) = <\boldsymbol{\rho}, e_S(P, \mathbf{O})>_{R^I} / <\boldsymbol{\rho}, \mathbf{1}>_{R^I} \quad (5)$$

where $e_S(P, \mathbf{O}) = \{e_S(P, \mathbf{O}_{N_i})\}_{i=1}^I$ and $\boldsymbol{\rho} = \{\rho_i\}_{i=1}^I$ is the vector of weighted coefficients defining priority of operations sequence $\mathbf{O}_{N_i}$, $i = \overline{1, I}$.

Of course, the quality of certain flow processing can not define quality of node operation in general. Quite a reasonable conclusion may be drawn on the basis of analysis of node operation regarding processing of collection of flows of different type that pass through the node over defined period of time. More detailed conclu-sions may be drawn while analysing node operation regarding carrying out operations sequences or separate operations with flows over similar time intervals.

Let us assume that flow movement in system is periodical and $T_0$ is minimal time interval that takes into account periodicity of flows passage through node $S$ that is defined by schedule. Let us suppose that $T^k$ is the period with duration of $T_0$, $\mathbf{P}^k = \{P_l^k\}_{l=1}^{L_k}$ is the collection of flows passing node $S$ over the $k$-th period, $k = \overline{1,K}$, $K$ is the number of test studies held over each flow from collection $\mathbf{P}^k$ during scheduled inspection of node operation. The quality $E_S^K(\mathbf{P}_l)$ of flow $P_l$ processing in node $S$ over $K$ periods $T_0$ is evaluated according to formula

$$E_S^K(\mathbf{P}_l) = <e_S(\mathbf{P}_l), \mathbf{1}>_{R^K} / K \qquad (6)$$

where $e_S(\mathbf{P}_l) = \{e_S(P_l^k)\}_{k=1}^{K}$ is the vector for evaluations of type (5) of flow $P_l \in \mathbf{P}_l = \{P_l^k\}_{k=1}^{K}$, $l = \overline{1,L_k}$, processing quality over $k$-th period $T_0$, $k = \overline{1,K}$.

Processing quality $E_S(\mathbf{P}^k)$ for collection of flows $\mathbf{P}^k$ that pass through the node $S$ over $k$-th period is determined according to formula

$$E_S(\mathbf{P}^k) = <\tilde{\boldsymbol{\rho}}, \mathbf{E}_S(\mathbf{P}^k)>_{R^{L_k}} / <\tilde{\boldsymbol{\rho}}, \mathbf{1}>_{R^{L_k}} \qquad (7)$$

where $\tilde{\boldsymbol{\rho}} = \{\rho_l\}_{l=1}^{L_k}$ is the vector of weighted coefficients defining priority of flows $P_l$, $l = \overline{1,L_k}$, $\mathbf{E}_S(\mathbf{P}^k) = \{E_S(P_l^k)\}_{l=1}^{L_k}$, $k = \overline{1,K}$. Similar generalized conclusions may be obtained for sequence of operations over the flows of separate types from collection $\mathbf{P}^k$ and, if needed, for particular operations with them.

The quality of node operation over $K$ periods $T_0$ with regard to condition of flow collection $\mathbf{P}^k$, $k = \overline{1,K}$, processing is determined according to formula

$$E_S^K(\mathbf{P}) = <\mathbf{E}_S(\mathbf{P}), \mathbf{1}>_{R^K} / K \qquad (8)$$

where $\mathbf{E}_S(\mathbf{P}) = \{E_S(\mathbf{P}^k)\}_{k=1}^{K}$. Analysis of sequence $E_S(\mathbf{P}^k)$, $k = \overline{1,K}$, allows to determine reasons for unsatisfactory processing of some categories of flows in the node.

Algorithm provided may be applied, for instance, for evaluation of trains of different categories processing quality at particular station of RTS. It often happens that results allow us to improve the quality of certain regular train processing by introducing minor changes into schedule. Here we have limited the procedure for local evaluation with the level of system's BSS, since while considering analysis of processing quality for flows that pass through the node, item-detail of process is usually excessive. Such a detailing is reasonable, in particular, while studying the process of train formation at marshalling yard etc.

We can not consider here all cases of behavior of characteristics of the elements of complex systems and corresponding methods of their local evaluation. Some additional examples can be found in [4, 6, 12, 14, 29].

## 5. Forecasting Evaluation

Forecasting evaluation may be short- and long-term and allows us to forecast both evaluations and system elements characteristics behaviour. Let us consider the algorithm for short-term forecasting of values of elements characteristic evaluations. Let us assume that $\{e(t_j)\}_{j=1}^{J}$, $J \geq 2$, is the prehistory of certain characteristic evaluations, obtained during performance of sequence of scheduled studies at the time points $t_j \in [0,T]$, $j = \overline{1,J}$, for time period $T$. Let us assume that $\boldsymbol{\Phi}(t) = \{\varphi_j(t)\}_{j=1}^{J}$ denotes the system of linearly independent functions determined on the interval $[0,T]$. Let us construct the function $e(t) = <\mathbf{A}, \boldsymbol{\Phi}(t)>_{R^J}$, where $\mathbf{A} = \{a_j\}_{j=1}^{J}$ is the vector of unknown coefficients. Then forecast value of evaluation $e(t)$ of selected characteristic at the time point $t_{J+1}$, for instance, at next scheduled study, is calculated from formula $e(t_{J+1}) = <\mathbf{A}, \boldsymbol{\Phi}(t_{J+1})>_{R^J}$, where the vector $\mathbf{A}$ is determined by the condition $<\mathbf{A}, \boldsymbol{\Phi}(t_k)>_{R^J} = e(t_k)$, $k = \overline{1,J}$. The selection of basis functions system may be determined by experimentally defined nature of behaviour characteristics evaluation of the object studied. Thus, the evaluation of state of RTS object is performed according to exponential law [28], which determines the selection of system $\boldsymbol{\Phi}(t)$. Outcomes of forecasting evaluation may be considered as the parts of local evaluation.

Here another drawback of integer rating evaluations is revealed, the point of which is the fact that they do not allow to obtain correct forecasts even for short-term prospect. Indeed, extrapolation of sequence $e(t_k)=3$, $k=1,2,3$, results in value $e(t_4)=3$. At the same time, sequence $e(t_1)=3.84$, $e(t_2) = 3.49$, $e(t_3)=3.11$ results in forecast $e(t_4) <3$, which is lower than "safety threshold". Forecasting analysis of adjusted evaluations allows us to define time point when conceptual evaluation will be decreased by unit under the same operation mode and absence of factors able to abruptly deteriorate system element's state or functioning quality. In particular, considering the behaviour of sequence $\{e(t_j)\}_{j=1}^{J}$, i.e. taking into account the fact that $e(t)$ is monotonically decreasing function, time for next inspection may be determined according to condition $e(t) \geq e^*$, where $e^*$ is the value that corresponds to conceptual evaluation, lower by unit than that one defined during the last study.

Term for short-term forecast $t_{J+1}$ is usually limited with the time of next scheduled evaluation of objects, the state of which improves abruptly as the result of repair following it. Procedure for forecasting element's behaviour evaluation consists in forecasting its values in points and is performed according to extrapolation algorithm described above. It is reasonable to perform such evaluation with regard to those characteristics only, for which the forecasting for the time point of next scheduled inspection was negative.

With regard to above formulated evaluation, it is obvious that forecasting analysis for pairs $e_c(f_{m,n,i})$, $e_l(f_{m,n,i})$ of characteristics $f_{m,n,i}$ behaviour allows to determine the evaluation of state of elementary section of track at the time point of its next scheduled inspection and to prevent it from crossing the "safety threshold", $i = \overline{1, I_{N_m}}$, $n = \overline{1, N_m}$, $m = \overline{1, M}$. Extrapolation of values of sequence $e_S(P_l^k)$ defined according to formula (5) allows to forecast the quality of flow $P_l, l = \overline{1, L_k}, k = \overline{1, K}$, processing in the node, for instance, processing of train at the station, for short-term prospects. The value of sequence $E_S(\mathbf{P}^k)$, $k = \overline{1, K}$, defined at (7) allows to perform forecasting analysis of system's node operation. For long-term forecasting, that include several periods of scheduled inspections and improvements following them we will use apparatus for time series analysis [30].

## 6. Aggregated Evaluation

Aggregated evaluation has been already used by us partially in formulas (1)-(4) when developing weighted average conclusion regarding state of BSS's element and in formulas (6)-(8) for developing aggregated evaluation of system node functioning quality. Generalization of evaluations of certain characteristics in uniform and mean-square metrics with regard to collection of elements composing BSS is performed consecutively according to formulas

$$V_{m,i}^c = <\tilde{\boldsymbol{\rho}}_m, \mathbf{e}^c(\tilde{\mathbf{f}}_{m,i})>_{R^{N_m}} / <\tilde{\boldsymbol{\rho}}_m, \mathbf{1}>_{R^{N_m}} \quad (9)$$

$$V_{m,i}^l = <\tilde{\boldsymbol{\rho}}_m, \mathbf{e}^l(\tilde{\mathbf{f}}_{m,i})>_{R^{N_m}} / <\tilde{\boldsymbol{\rho}}_m, \mathbf{1}>_{R^{N_m}} \quad (10)$$

where $\mathbf{e}^c(\tilde{\mathbf{f}}_{m,i}) = \{e^c(f_{m,n,i})\}_{n=1}^{N_m}$, $\mathbf{e}^l(\tilde{\mathbf{f}}_{m,i}) = \{e^l(f_{m,n,i})\}_{n=1}^{N_m}$ are the vectors of adjusted rating evaluations for characteristic $f_{m,n,i}$ for collection of elements of BSS $D_m$ in uniform and mean-square metrics, $\tilde{\boldsymbol{\rho}}_m = \{\tilde{\rho}_{m,n}\}_{n=1}^{N_m}$ is the vector of weighted coefficients defining the priority of elements in BSS structure. Then, generalized conclusion regarding state of elementary section $D_{m,n}$ for characteristic determined is defined according to formula

$$V_{m,i} = (\rho^c V_{m,i}^c + \rho^l V_{m,i}^l)/(\rho^c + \rho^l), \; i = \overline{1, I_{N_m}}, \quad (11)$$

As it is obvious from formulas (1)-(3) and (9)-(11), we develop aggregated evaluations in two directions, which we shall arbitrarily call "horizontal" (H) and "vertical" (V). The first one develops aggregated conclusions regarding state or functioning quality for separate BSS elements with regard to set of parameters, criteria and characteristics subject to evaluation. It allows us to identify elements that has obtained unsatisfactory or close to unsatisfactory evaluations. The second direction forms aggregated conclusions regarding state or operating quality of BSS with regard to separate parameter, criterion or characteristic of elements that compose BSS. It allows us to determine parameters, criteria or characteristics regarding to which BSS evaluations are also unsatisfactory or close to unsatisfactory. Regarding RTS objects, "horizontal" direction of track evaluation allows us to discover "critical" sections of certain inter-station railway track that require urgent repair. "Vertical" direction allows us to discover problem-causing characteristics, for example, roadbed pollution, state of rails or ties on inter-station railway track etc.

The level of evaluation coverage of BSS $D_{m,n}$ elements with consideration of their priority is defined according to formula

$$C_{D_m} = <\tilde{\boldsymbol{\rho}}_m, \mathbf{C}_{D_m}>_{R^{N_m}} / <\tilde{\boldsymbol{\rho}}_m, \mathbf{1}>_{R^{N_m}} \quad (12)$$

where $\mathbf{C}_{D_m} = \{C_{D_{m,n}}\}_{n=1}^{N_m}$.

Generalized conclusion regarding state of BSS $D_m$ at the time point of last scheduled inspection with regard to collection of elements and their characteristics, obtained as the result of scheduled study is defined according to formula

$$E_{D_m}^{s,P} = <\tilde{\boldsymbol{\rho}}_m, \mathbf{H}_m>_{R^{N_m}} / <\tilde{\boldsymbol{\rho}}_m, \mathbf{1}>_{R^{N_m}} \equiv \quad (13)$$

$$\equiv <\boldsymbol{\rho}_m, e_c(f_{m,n,i})>/<\boldsymbol{\rho}_m, \mathbf{1}>_{R^{I_{N_m}}}$$

where $\mathbf{H}_m = \{H_{m,n}\}_{n=1}^{N_m}$, $\mathbf{V}_m = \{V_{m,i}\}_{i=1}^{I_{N_m}}$.

Same method is applied to developing local forecasting and weighted average evaluations of state of nodes $S_m$, $m = \overline{1, \tilde{M}}$, and functioning quality of edges and nodes of subsystem SD at the time points of their last and next scheduled studies.

The next evaluation level consists in development of sequence of weighted aggregated evaluations that form the conclusion regarding state and functioning quality of subsystem SD in general. In particular, generalized conclusion regarding BSS $D_m$ at the time points of last and next scheduled inspections is defined according to formulas

$$H_{D_m}^P = (\rho^{s,P} E_{D_m}^{s,P} + \rho^{f,P} E_{D_m}^{f,P})/(\rho^{s,P} + \rho^{f,P}) \quad (14)$$

$$H_{D_m}^F = (\rho^{s,F} E_{D_m}^{s,F} + \rho^{f,F} E_{D_m}^{f,F})/(\rho^{s,F} + \rho^{f,F}) \quad (15)$$

$$H_{D_m} = (\rho^P H_{D_m}^P + \rho^F H_{D_m}^F)/(\rho^P + \rho^F) \quad (16)$$

where $E_{D_m}^{s,P}$, $E_{D_m}^{f,P}$, $E_{D_m}^{s,F}$, and $E_{D_m}^{f,F}$ are aggregated evaluations of state and quality of BSS $D_m$ functioning at the present and following terms of scheduled studies accordingly, $\rho^{s,P}$, $\rho^{f,P}$, $\rho^{s,F}$, and $\rho^{f,F}$ are weighted coefficients defining their priority, $\rho^P$ and $\rho^F$ are weighted coefficients defining priority of present and forecast evaluations $H_{D_m}^P$ and $H_{D_m}^F$, respectively, (in formulas (14)-(16) and below the following indices are used: $s$ – state, $f$ – function, $P$ – Present, $F$ – Forecast). Similarly are generalized present and forecasting conclusions $H_{S_m}^P$, $H_{S_m}^F$, and $H_{S_m}$ regarding subsystem nodes made $SD$.

Aggregated conclusion regarding state and functioning quality of all BSSs of subsystem $SD$ at the time point of the last and following scheduled inspections is made consecutively as follows:

– state of nodes and edges at the time point of last scheduled study is defined according to formula

$$V_{SD}^{s,P} = <\boldsymbol{\rho}_D, \boldsymbol{E}_{\mathbf{D}}^{s,P}>_{R^M} / <\boldsymbol{\rho}_D, \mathbf{1}>_{R^M} +$$
$$+ <\boldsymbol{\rho}_S, \boldsymbol{E}_{\mathbf{S}}^{s,P}>_{R^{\tilde{M}}} / <\boldsymbol{\rho}_S, \mathbf{1}>_{R^{\tilde{M}}} \quad (17)$$

– state of nodes and edges at the time point of following scheduled study is defined according to formula

$$V_{SD}^{s,F} = <\boldsymbol{\rho}_D, \boldsymbol{E}_{\mathbf{D}}^{s,F}>_{R^M} / <\boldsymbol{\rho}_D, \mathbf{1}>_{R^M} +$$
$$+ <\boldsymbol{\rho}_S, \boldsymbol{E}_{\mathbf{S}}^{s,F}>_{R^{\tilde{M}}} / <\boldsymbol{\rho}_S, \mathbf{1}>_{R^{\tilde{M}}} \quad (18)$$

– quality of nodes and edges functioning at the time point of last scheduled study is defined according to formula

$$V_{SD}^{f,P} = <\boldsymbol{\rho}_D, \boldsymbol{E}_{\mathbf{D}}^{f,P}>_{R^M} / <\boldsymbol{\rho}_D, \mathbf{1}>_{R^M} +$$
$$+ <\boldsymbol{\rho}_S, \boldsymbol{E}_{\mathbf{S}}^{f,P}>_{R^{\tilde{M}}} / <\boldsymbol{\rho}_S, \mathbf{1}>_{R^{\tilde{M}}} \quad (19)$$

– quality of nodes and edges functioning at the time point of following scheduled study is defined according to formula

$$V_{SD}^{f,F} = <\boldsymbol{\rho}_D, \boldsymbol{E}_{\mathbf{D}}^{f,F}>_{R^M} / <\boldsymbol{\rho}_D, \mathbf{1}>_{R^M} +$$
$$+ <\boldsymbol{\rho}_S, \boldsymbol{E}_{\mathbf{S}}^{f,F}>_{R^{\tilde{M}}} / <\boldsymbol{\rho}_S, \mathbf{1}>_{R^{\tilde{M}}} \quad (20)$$

In formulas (17)-(20) $\boldsymbol{E}_{\mathbf{D}}^{s,P} = \{E_{D_m}^{s,P}\}_{m=1}^M$, $\boldsymbol{E}_{\mathbf{D}}^{s,F} = \{E_{D_m}^{s,F}\}_{m=1}^M$, $\boldsymbol{E}_{\mathbf{S}}^{s,P} = \{E_{S_m}^{s,P}\}_{m=1}^{\tilde{M}}$, $\boldsymbol{E}_{\mathbf{S}}^{s,F} = \{E_{S_m}^{s,F}\}_{m=1}^{\tilde{M}}$, and $\boldsymbol{\rho}_D = \{\rho_{D_m}\}_{m=1}^M$, $\boldsymbol{\rho}_S = \{\rho_{S_m}\}_{m=1}^{\tilde{M}}$ are the vectors of weighted coefficients that define priority of separate edges and nodes of subsystem $SD$.

State and functioning quality of all BSSs of system $SD$ considering forecast are defined according to formulas

$$V_{SD}^s = (\rho^P V_{SD}^{s,P} + \rho^F V_{SD}^{s,F})/(\rho^P + \rho^F) \quad (21)$$
$$V_{SD}^f = (\rho^P V_{SD}^{f,P} + \rho^F V_{SD}^{f,F})/(\rho^P + \rho^F) \quad (22)$$

Generalized level of evaluation coverage of characteristics of subsystem's $SD$ BSS elements are defined according to formula

$$C_{SD} = <\boldsymbol{\rho}_D, C_\mathbf{D}>_{R^M} / <\boldsymbol{\rho}_D, \mathbf{1}>_{R^M} +$$
$$+ <\boldsymbol{\rho}_S, C_\mathbf{S}>_{R^{\tilde{M}}} / <\boldsymbol{\rho}_S, \mathbf{1}>_{R^{\tilde{M}}} \quad (23)$$

where $C_\mathbf{D} = \{C_{D_m}\}_{m=1}^M$, $C_\mathbf{S} = \{C_{S_m}\}_{m=1}^{\tilde{M}}$.

Final conclusion regarding subsystem $SD$ quality is defined from formula

$$E_{SD} = (\rho^s V_{SD}^s + \rho^f V_{SD}^f)/(\rho^s + \rho^f) \equiv$$
$$\equiv <\boldsymbol{\rho}_D, H_\mathbf{D}>_{R^M} / <\boldsymbol{\rho}_D, \mathbf{1}>_{R^M} + \quad (24)$$
$$+ <\boldsymbol{\rho}_S, H_\mathbf{S}>_{R^{\tilde{M}}} / <\boldsymbol{\rho}_S, \mathbf{1}>_{R^{\tilde{M}}}$$

where $H_\mathbf{D} = \{H_{D_m}\}_{m=1}^M$, $H_\mathbf{S} = \{H_{S_m}\}_{m=1}^{\tilde{M}}$, and $\rho^s$, $\rho^f$ are weighted coefficients defining priority of evaluations of state and functioning quality of subsystem $SD$.

Similarly are aggregated evaluations for subsystems of all hierarchy levels developed. Even for those systems like RTS, despite the huge general scopes of evaluation results number of which may comprise millions at local level, maximal number of localization steps in proposed methods would not exceed seven (RTS → territorial railway → directory → distance → division → BSS → element → → characteristic).

## 7. Interactive Evaluation

Interactive evaluation will be performed on the level of interaction analysis for objects like flow $P_j$, $j = \overline{1,M}$, and line $[S_1, S_{N+1}]$ (sequence of nodes $S_i$, $i = \overline{1, N+1}$, and edges $D_i = (S_i, S_{i+1})$, $i = \overline{1, N}$, connecting them). Here $M$ is the number of flows that pass through the line $[S_0, S_N]$ over certain time interval. Let us assume that flows passage through the line is completely determined, i. e. schedule of their movement is completely defined.

Flow delay over the edge may be caused by circumstances like unsatisfactory state of edge, unsatisfactory state of flow, node being not ready to accept the flow etc. Out of all these reasons only the first one is in direct relation with state of the edge. Flow delay in node may be caused by circumstances like unsatisfactory state or organization of node's functioning, unsatisfactory state of flow, inability to dispatch flow because of next edge on movement direction being occupied by another flows etc. Similarly to previous case, out of all these reasons only the first one is in direct relation to organization of node's functioning. With the passage of flows through the line, effect from factors given may be consecutively accumulated or compensated. Part of them is of random nature, the rest may be regular. The main purpose of interactive evaluation is discovering and localization of regular negative factors that stipulate deviation from established schedule of flows movement.

Let us denote the time for flow processing $P_j$ in node $S_i$ according to schedule with $t_{j,i}^{n,s}$, minimum possible time for its processing in given node – with $t_{j,i}^{n,\min}$, real time for flow processing – with $t_{j,i}^{n,r}$, time for which the flow $P_j$ passes through the edge $D_i$ according to schedule – with $t_{j,i}^{e,s}$, minimum possible time for its passage – with $t_{j,i}^{e,\min}$, real time for which the flow passes through the edge – with $t_{j,i}^{p,r}$, $j=\overline{1,M}$, $i=\overline{1,N}$. Let us denote minimum time interval which takes into account periodicity of flow movement determined by schedule with $T^0$. Let $T_k$ be the period of duration $T^0$ with running number $k$, $k=\overline{1,K}$, $T^K = KT^0$. Usually the value $T^K$ does not exceed duration of time interval between scheduled inspections of objects composing the line. We shall consider that evaluation $e(P_j, S_i, T_k)$ of flow $P_j$ processing quality in node $S_i$ for period $T_k$ is equal to:

– 5, if $t_{j,i}^{n,r} = t_{j,i}^{n,\min}$, i. e. processing time compensate previous delays in movement to maximal extent;

– 4+ $(t_{j,i}^{n,r} - t_{j,i}^{n,s})/(t_{j,i}^{n,\min} - t_{j,i}^{n,s})$, if $t_{j,i}^{n,r} \in (t_{j,i}^{n,\min}, t_{j,i}^{n,s}]$, i. e. processing time partially compensate previous delays in movement;

– 3+$(t_{j,i}^{n,s} + t_{j,i}^{e,s} - t_{j,i}^{e,\min} - t_{j,i}^{n,r})/(t_{j,i}^{e,s} - t_{j,i}^{e,\min})$, if $t_{j,i}^{n,r} \in (t_{j,i}^{n,s}, t_{j,i}^{n,s} + (t_{j,i}^{e,s} - t_{j,i}^{e,\min})]$, i.e. flow delay in the node may be fully compensated on the following edge, for example, due to safe increase in flow movement speed;

– 2, if $t_{j,i}^{n,r} < t_{j,i}^{n,s} + (t_{j,i}^{e,s} - t_{j,i}^{e,\min})$, i.e. flow delay in the node can not be compensated on the following edge, $j=\overline{1,M}$, $i=\overline{1,N}$, $k=\overline{1,K}$.

We shall consider that evaluation $e(P_j, D_i, T_k)$ of quality of flow $P_j$ passage through the edge $D_i$ for period $T_k$ is equal to:

– 5, if $t_{j,i}^{e,r} = t_{j,i}^{e,\min}$, i.e. time for flow passing through the edge compensates previous delays in movement to maximal extent;

– 4+$(t_{j,i}^{e,r} - t_{j,i}^{e,s})/(t_{j,i}^{e,\min} - t_{j,i}^{e,s})$, if $t_{j,i}^{e,r} \in (t_{j,i}^{e,\min}, t_{j,i}^{e,s}]$, i.e. time for flow passing through the edge partially compensate previous delays in movement;

– 3+$(t_{j,i}^{e,s} + t_{j,i}^{n,s} - t_{j,i}^{n,\min} - t_{j,i}^{e,r})/(t_{j,i}^{n,e} - t_{j,i}^{n,\min})$, if $t_{j,i}^{e,r} \in (t_{j,i}^{e,s}, t_{j,i}^{e,s} + (t_{j,i}^{n,s} - t_{j,i}^{n,\min})]$, i.e. flow delay on the edge may be fully compensated due to decrease in processing time in following node;

– 2, if $t_{j,i}^{e,r} < t_{j,i}^{e,g} + (t_{j,i}^{n,s} - t_{j,i}^{n,\min})$, i.e. flow delay on the edge can not be compensated in following node, $j=\overline{1,M}$, $i=\overline{1,N}$, $k=\overline{1,K}$.

Necessity of compensation of previous delays increases intensity of operation of node or flow support on the edge, and related risks as well. Numerically, number of compensations and efforts for their realization is implemented by fractional parts of evaluations $e(P_j, S_i, T_k)$, that exceed the value of 4, or fractional parts of evaluations $e(P_j, D_{i-1}, T_k)$ that are smaller than this value. Thus, evaluation $e(P_j, S_i, T_k) = 4,68$ means that flow $P_j$ support operation works over time interval $T_k$ were quite effective, as long as all the operations with flow in node were carried out for the time specified, moreover, all previous delays were compensated. However, quite large compensation (0.68) shows that delays were quite significant. If the value $e(P_j, S_i, T_k) < 4$, delay compensation is also defined by fractional part of evaluation and may be implemented on the following edge on the way by means of flow speed increase.

It is obvious that single evaluation of separate flow delay in certain node or on certain edge, for example train delay on the station or inter-station railway track, can not be final factor defining their state or functioning quality. More reasonable conclusion can be made if the delays are evaluated for flow or several flows passing through separate node or sequence of nodes and edges on the line during specified time period $T^K$. Such evaluations allow us to localize, at least partially, reasons for drawbacks in functioning of separate CHNS objects located on one line or passing through it.

Evaluation of node $S_i$ and edge $D_i$ with regard to results of flow $P_j$ processing over the period $T^K$ is defined according to formulas

$$E(P_j, S_i, T^K) = <\mathbf{1}, \mathbf{e}(P_j, S_i, \mathbf{T})>_{R^K} / K \quad (25)$$

$$E(P_j, D_i, T^K) = <\mathbf{1}, \mathbf{e}(P_j, D_i, \mathbf{T})>_{R^K} / K \quad (26)$$

Here $\mathbf{e}(P_j, S_i, \mathbf{T}) = \{e(P_j, S_i, T_k)\}_{k=1}^K$, $\mathbf{e}(P_j, D_i, \mathbf{T}) = \{e(P_j, D_i, T_k)\}_{k=1}^K$, and $\mathbf{T} = \{T_k\}_{k=1}^K$. With consecutive increase of interval $T^K$, values $E(P_j, S_i, T^K)$ allow to follow up the dynamics of changes in flow $P_j$ processing quality in the node $S_i$, and $E(P_j, D_i, T^K)$ – that one of its passing over the edge $D_i$, $j=\overline{1,M}$, $i=\overline{1,N}$.

g ith regard to r

$\}_j^M$ over th

period ned according to formulas

$$E_{\mathbf{P}}(S_i, T_k) = <\mathbf{R_P}, \mathbf{e}(\mathbf{P}, S_i, T_k)>_{R^M} / <\mathbf{R_P}, \mathbf{1}>_{R^M} \quad (27)$$

$$E_{\mathbf{P}}(D_i, T_k) = <\mathbf{R_P}, \mathbf{e}(\mathbf{P}, D_i, T_k)>_{R^M} / <\mathbf{R_P}, \mathbf{1}>_{R^M} \quad (28)$$

where $\mathbf{e}(\mathbf{P}, S_i, T_k) = \{e(P_j, S_i, T_k)\}_{j=1}^M$, $\mathbf{e}(\mathbf{P}, D_i, T_k) = \{e(P_j, D_i, T_k)\}_{j=1}^M$, and $\mathbf{R_P} = \{\rho_{P_j}\}_{j=1}^M$ is the vector of weighted coefficients defining priority of flows in collection $\{P_j\}_{j=1}^M$. After calculating $E_{\mathbf{P}}(S_i, T_k)$ and $E_{\mathbf{P}}(D_i, T_k)$ for each of $k$ periods the sequences of evaluations will be obtained, the analysis of which allows to detect cyclic changes and forecast the behaviour of flows processing quality in separate node or their passage through edge of the line.

Evaluation of node $S_i$ and edge $D_i$ with regard to results of passage of flows collection $\mathbf{P}$ over the period $T^K$ is defined according to formulas

$$E_{\mathbf{P}}(S_i, T^K) = <\mathbf{1}, \mathbf{E_P}(S_i, \mathbf{T})>_{R^K} / K \quad (29)$$

$$E_{\mathbf{P}}(D_i, T^K) = <\mathbf{1}, \mathbf{E_P}(D_i, \mathbf{T})>_{R^K} / K \quad (30)$$

where $\mathbf{E_P}(S_i, \mathbf{T}) = \{E_{\mathbf{P}}(S_i, T_k)\}_{k=1}^K$ and $\mathbf{E_P}(D_i, \mathbf{T}) = \{E_{\mathbf{P}}(D_i, T_k)\}_{k=1}^K$. With consequent increase $T^K$ values of those evaluations allow to follow up the dynamics of changes in flows processing quality in the node $S_i$ and their passing over the edges $D_i$, $i = \overline{1, N}$. If the evaluation $E(P_j, S_i, T^K)$ of flow $P_j$ processing in the node $S_i$ over the period $T^K$ is significantly smaller than $E_{\mathbf{P}}(S_i, T^K)$ and/or aggregated evaluation $E(P_j, D_i, T^K)$ of its passage over the edge $D_i$ over the period $T^K$ is significantly smaller than $E_{\mathbf{P}}(D_i, T^K)$, the conclusion can be made regarding presence of drawbacks in schedule of this flow movement.

Evaluations of flow $P_j$ processing in sequence of nodes $\mathbf{S} = \{S_i\}_{i=1}^N$ and its passage through sequence of edges $\mathbf{D} = \{D_i\}_{i=1}^N$ located on the line $[S_0, S_N]$ over the period of time $T_k$, are defined according to formulas

$$E_{\mathbf{S}}(P_j, T_k) = <\mathbf{R_S}, \mathbf{e}(P_j, \mathbf{S}, T_k)>_{R^N} / <\mathbf{R_S}, \mathbf{1}>_{R^N} \quad (31)$$

$$E_{\mathbf{D}}(P_j, T_k) = <\mathbf{R_D}, \mathbf{e}(P_j, \mathbf{D}, T_k)>_{R^N} / <\mathbf{R_D}, \mathbf{1}>_{R^N} \quad (32)$$

where $\mathbf{e}(P_j, \mathbf{S}, T_k) = \{e(P_j, S_i, T_k)\}_{i=1}^N$, $\mathbf{e}(P_j, \mathbf{D}, T_k) = \{e(P_j, D_i, T_k)\}_{i=1}^N$, and $\mathbf{R_S} = \{\rho_{S_i}\}_{i=1}^N$, $\mathbf{R_D} = \{\rho_{D_i}\}_{i=1}^N$ are the vectors of weighted coefficients defining the priority of nodes and edges over the line $[S_0, S_N]$ respectively. Analysis of sequences $E_{\mathbf{S}}(P_j, T_k)$ and $E_{\mathbf{D}}(P_j, T_k)$, $j = \overline{1, M}$, $k = \overline{1, K}$, allows to discover cyclic changes and to perform the forecasting of quality of flow $P_j$ processing in the nodes and its passage through the edges of the line $[S_0, S_N]$.

Evaluations of flow $P_j$ processing in sequence of nodes $\{S_i\}_{i=1}^N$ and its passage through sequence of edges $\{D_i\}_{i=1}^N$ located on the line $[S_0, S_N]$ over the time interval $T^K$, are defined according to formulas

$$E_{\mathbf{S}}(P_j, T^K) = <\mathbf{1}, \mathbf{E_S}(P_j, \mathbf{T})>_{R^K} / K \quad (33)$$

$$E_{\mathbf{D}}(P_j, T^K) = <\mathbf{1}, \mathbf{E_D}(P_j, \mathbf{T})>_{R^K} / K \quad (34)$$

respectively, where $\mathbf{E_S}(P_j, \mathbf{T}) = \{E_{\mathbf{S}}(P_j, T_k)\}_{k=1}^K$ and $\mathbf{E_D}(P_j, \mathbf{T}) = \{E_{\mathbf{D}}(P_j, T_k)\}_{k=1}^K$. With consequent increase $T^K$ values $E_{\mathbf{S}}(P_j, T^K)$ and $E_{\mathbf{D}}(P_j, T^K)$, $j = \overline{1, M}$, allow to follow up the dynamics of changes in flow $P_j$ processing quality in the nodes and its passing over the edges $[S_0, S_N]$.

Regarding the objects of RTS, the evaluations (25)-(26) allow to analyse dynamics of changes in quality of particular train processing on the station or its passing over inter-station railway track, evaluations (27)-(30) – to follow up the dynamics of changes in quality of train sequence processing on separate station or their passage over separate inter-station railway track within determined unit of time or larger time interval, as well as to discover the drawbacks in established schedule of trains movement, evaluations (31)-(34) – to analyse the quality of processing of particular train on the sequence of stations or its passing through inter-station railway tracks on the line and schedule's sensitivity to minor delays in movement. Similar examples of application of interactive evaluation results may be provided for many more real CHNS, in particular, with partially determined movement of flows over the network.

Evaluation of flows collection $\{P_j\}_{j=1}^M$ processing in the sequence of nodes $\{S_i\}_{i=1}^N$ located on the line $[S_0, S_N]$ over the $T_k$ period of time is defined according to formula

$$E_{\mathbf{P,S}}(T_k) = <\mathbf{R_P}, \mathbf{E_S}(\mathbf{P}, T_k)>_{R^M} / <\mathbf{R_P}, \mathbf{1}>_{R^M} =$$

$$= <\mathbf{R_S}, \mathbf{E_P}(\mathbf{S}, T_k)>_{R^N} / <\mathbf{R_S}, \mathbf{1}>_{R^N} \quad (35)$$

where $\mathbf{E_S}(\mathbf{P}, T_k) = \{E_{\mathbf{S}}(P_j, T_k)\}_{j=1}^M$ and $\mathbf{E_P}(\mathbf{S}, T_k) = \{E_{\mathbf{P}}(S_i, T_k)\}_{i=1}^N$, $k = \overline{1, K}$. Analysis of this sequence allows to discover cyclic changes in processing quality of all flows passing through the nodes on the line $[S_0, S_N]$, within interval $T_k$.

Evaluation of flows collection $\{P_j\}_{j=1}^M$ passing through the sequence of edges $\{D_i\}_{i=1}^N$ located on the line $[S_0, S_N]$ over the $T_k$ period of time is defined according to formula

$$E_{\mathbf{P,D}}(T_k) = <\mathbf{R_P}, \mathbf{E_D}(\mathbf{P}, T_k)>_{R^M} / <\mathbf{R_P}, \mathbf{1}>_{R^M} =$$
$$= <\mathbf{R_D}, \mathbf{E_P}(\mathbf{D}, T_k)>_{R^N} / <\mathbf{R_D}, \mathbf{1}>_{R^N} \quad (36)$$

where $\mathbf{E_D}(\mathbf{P}, T_k) = \{E_\mathbf{D}(P_j, T_k)\}_{j=1}^M$ and $\mathbf{E_P}(\mathbf{D}, T_k) = \{E_\mathbf{P}(D_i, T_k)\}_{i=1}^N$, $k = \overline{1, K}$. Analysis of this sequence allows to discover cyclic changes in quality of all flows passing over the edges on the line $[S_0, S_N]$, within interval $T_k$.

Evaluations of flows collection $\{P_j\}_{j=1}^M$ processing in sequence of nodes $\{S_i\}_{i=1}^N$ and their passage through sequence of edges $\{D_i\}_{i=1}^N$ located on the line $[S_0, S_N]$ over time interval $T^K$, are defined according to formulas

$$E_{\mathbf{P,S}}(T^K) = <\mathbf{1}, \mathbf{E_{P,S}(T)}>_{R^K} / K \quad (37)$$

$$E_{\mathbf{P,D}}(T^K) = <\mathbf{1}, \mathbf{E_{P,D}(T)}>_{R^K} / K \quad (38)$$

where $\mathbf{E_{P,S}(T)} = \{E_{\mathbf{P,S}}(T_k)\}_{k=1}^K$ and $\mathbf{E_{P,D}(T)} = \{E_{\mathbf{P,D}}(T_k)\}_{k=1}^K$. With consecutive growth $T^K$, values $E_{\mathbf{P,S}}(T^K)$ and $E_{\mathbf{P,D}}(T^K)$ allow to follow up the dynamics of changes in quality of flows collection $\{P_j\}_{j=1}^M$ processing in the nodes and passing through the edges located on the line $[S_0, S_N]$. If generalized evaluation $E_\mathbf{P}(S_i, T^K)$ of node $S_i$ over the time period $T^K$ is significantly smaller than $E_{\mathbf{P,S}}(T^K)$, reasonable conclusion can be made regarding presence of significant drawbacks in state of its infrastructure and functioning organization. Similarly, if generalized evaluation $E_\mathbf{P}(D_i, T^K)$ of the edge $D_i$ over the time period $T^K$ is significantly smaller than $E_{\mathbf{P,D}}(T^K)$, reasonable conclusion can be made regarding presence of significant drawbacks in its state. Such conclusions are strong reason for out-of-schedule inspection of state or functioning quality of corresponding system's objects.

Above, when generalizing evaluations for nodes and edges of network structure, we have been separating them as for the objects of CHNSs of different types. Aggregated evaluation of flows collection $\{P_j\}_{j=1}^M$ passing through the line $[S_0, S_N]$ in general over the period $T_k$ is defined according to formula

$$E_\mathbf{P}(T_k) = (\rho_\mathbf{S} E_{\mathbf{P,S}}(T_k) + \rho_\mathbf{D} E_{\mathbf{P,D}}(T_k))/(\rho_\mathbf{S} + \rho_\mathbf{D}) \quad (39)$$

where $E_{\mathbf{P,S}}(T_k)$ and $E_{\mathbf{P,D}}(T_k), k = \overline{1, K}$, defined according to (35), (36) and $\rho_\mathbf{S}$, $\rho_\mathbf{D}$ are weighted coefficients defining priority of collections of nodes and edges that compose the line, during the evaluation process. Thus, after the repair of track and corresponding stabilization period, greater attention is normally paid to state and functioning effectiveness of stations. Analysis of the last sequence allows to detect cyclic changes in quality of flows collection $\{P_j\}_{j=1}^M$ processing on the line $[S_0, S_N]$ in general.

Average evaluation of collection of flows $\{P_j\}_{j=1}^M$ passing through the line $[S_0, S_N]$ over time interval $T^K$ is defined according to formula

$$E_\mathbf{P}(T^K) = <\mathbf{1}, \mathbf{E_P(T)}>_{R^K} / K \quad (40)$$

where $\mathbf{E_P(T)} = \{E_\mathbf{P}(T_k)\}_{k=1}^K$. With consecutive growth $T^K$, values $E_\mathbf{P}(T^K)$ allow to follow up the dynamics of changes in quality of flows collection $\{P_j\}_{j=1}^M$ processing on the line $[S_0, S_N]$. If generalized evaluation of separate flow $E(P_j, T^K)$ is much lower than $E_\mathbf{P}(T^K)$, the reasonable conclusion can be made regarding its possible unsatisfactory state or necessity in change of movement schedule.

Regarding the objects of RTS, evaluations (35)-(38) allow to make generalized conclusions regarding collection of trains passing stations and inter-station railway tracks of the line over determined period or long intervals of time, which allows to discover railway objects, functioning of which is unsatisfactory, evaluations (39)-(40) – to analyse state and effectiveness of organization of train movement and line in general.

If, for the collection of flows that pass through the line over time period $T^K$, delays are normally compensated in nodes, this is indirect, though quite indicative, factor defining the quality of flow or edge state. On the other hand, if delays are compensated on the edges, this indicates infrastructure quality or nodes functioning effectiveness. If generalized evaluation of compensations, which is indicator of their mass character is lower than aggregated evaluation of CHNS's objects forming up a line, the conclusion can be made that flow movement schedule established on this line is not optimal and is sensitive to minor delays. In general, when following up the dynamics of aggregated interactive evaluations of all levels with consecutive growth of value $T^K$, we can define trends in changes of state and functioning quality of corresponding CHNS's objects. Meanwhile, short-term forecasting obtained as the result of interactive evaluations extrapolation on the basis of known prehistory of their values allows us to discover beforehand objects, which, in the closest prospect, may cross the "safety threshold", i.e. they require out-of-schedule inspection and corresponding measures taken. Long-term forecasting of interactive evaluations performed, for example, by means of apparatus for time series analysis, allows to follow up season changes in behaviour of main CHNS structural elements and to prevent negative trends of their development.

# 8. Conclusions

This work proposes methodology for complex evaluation of real large complex dynamical system with hierarchically-network structure, the component of which are the methods for local, forecasting, interactive and aggregated evaluation of its main objects. It is showed that when combined, together with use of adjusted rating scale they form up quite comprehensive, adequate, and integral notion regarding state, functioning quality and interaction of objects of studied system and its subsystems on all levels of its structure. Described methodology is applied for development of software for evaluation of state and functioning quality of track and station facilities of Ukrainian Railways [12, 20]. Separate methods of proposed methodology were applied for evaluation of quality of prosthesis of lower limbs of disabled persons and level of recovery of functional capabilities of human musculoskeletal system at different cases of pathologies and means of rehabilitation [11, 29] which indicates universality of proposed approach for complex systems evaluation. Results of the evaluations should be used to make decision about how to proceed with regard to the objects of estimation [31].